\include{graphicsx} 
\documentclass[12pt,preprint]{aastex}
\usepackage{multirow}
\newcommand{\Lsol}{\mbox{$L_\odot$}}
\newcommand{\Msol}{\mbox{$M_\odot$}}

\slugcomment{}
\shorttitle{New Extinction and Mass Estimates of CT Cha B with MagAO}
\shortauthors{Wu et al.}

\begin{document}

\title{New Extinction and Mass Estimates from Optical Photometry of the Very Low Mass Brown Dwarf Companion CT Chamaeleontis B with the Magellan AO System\footnote{This paper includes data gathered with the 6.5 m Magellan Clay Telescope at Las Campanas Observatory, Chile.}} 

\author{Ya-Lin Wu$^1$, 
 Laird M. Close$^1$, 
 Jared R. Males$^{1}\footnote{NASA Sagan Fellow}$ ,
 Travis S. Barman$^2$, 
 Katie M. Morzinski$^{1}\footnotemark[2]$ , 
 Katherine B. Follette$^1$,
 Vanessa Bailey$^1$,
 Timothy J. Rodigas$^{1,3}\footnote{Carnegie Postdoctoral Fellow}$ , 
 Philip Hinz$^1$,
 Alfio Puglisi$^4$, 
 Marco Xompero$^4$,
 Runa Briguglio$^4$}

\email{yalinwu@email.arizona.edu}

\affil{$^1$Steward Observatory, University of Arizona, Tucson, Arizona 85721, USA}
\affil{$^2$Lunar and Planetary Laboratory, University of Arizona, Tucson, Arizona 85721, USA}
\affil{$^3$Department of Terrestrial Magnetism, Carnegie Institute of Washington, 5241 Broad
Branch Road, NW, Washington, DC 20015, USA}
\affil{$^4$INAF - Osservatorio Astrofisico di Arcetri, Largo E. Fermi 5, I-50125, Firenze, Italy}

\begin{abstract} 
We used the Magellan adaptive optics (MagAO) system and its VisAO CCD camera to image the young low mass brown dwarf companion CT Chamaeleontis B for the first time at visible wavelengths. We detect it at $r'$, $i'$, $z'$, and $Y_S$. With our new photometry and $T_{eff} \sim2500$ K derived from the shape its $K-$band spectrum, we find that CT Cha B has $A_V=3.4\pm1.1$ mag, and a mass of $14-24$ $M_J$ according to the DUSTY evolutionary tracks and its $1-5$ Myr age. The overluminosity of our $r'$ detection indicates that the companion has significant H$\alpha$ emission and a mass accretion rate $\sim6\times10^{-10 } \Msol/\mbox{yr}$, similar to some substellar companions. Proper motion analysis shows that another point source within $2\arcsec$ of CT Cha A is not physical. This paper demonstrates how visible wavelength AO photometry ($r'$, $i'$, $z'$, $Y_S$) allows for a better estimate of extinction, luminosity, and mass accretion rate of young substellar companions.
\end{abstract}


\section{INTRODUCTION}
As more and more brown dwarfs and planetary companions are being discovered, characterizing them in the visible regime yields a more complete picture of the spectral energy distribution (SED) and more insight into physical properties as well as formation scenarios. For instance, a better estimate of extinction helps to derive bolometric luminosity and mass---especially for young objects ($\leq10$ Myr) which may still have primeval dust and gas around them and suffer significant obscuration. However, extinction is problematic to measure because most of the high-contrast adaptive optics (AO) observations are done in the near-infrared, which is $\sim10$ times less sensitive to dust at $K$ versus $V$. One simple treatment is to assume that the companion has the same amount of extinction as its host star \citep{P12}, since in the early stages of star formation the binary might be embedded in a common envelope. For more evolved, fragmented systems both components may have their own disks, so this assumption might be invalid. Ideally one would like to acquire visible spectra or at least broad-band visible photometry to supplement near-IR measurements because visible wavelengths are a better probe for dust extinction. Yet high contrast optical observations on companions are very rare due to decreased contrast \citep{M14} and the difficulty of correcting atmospheric turbulence at visible (defined here as $\lambda<1\mu$m) wavelengths. We therefore need an advanced AO system which can work in the visible to suppress the halo.

Here we present the first optical AO photometry of CT Chamaeleontis system with the Magellan adaptive optics (MagAO) system, a powerful new 585-element AO system commissioned on the 6.5 m Clay Telescope \citep{C13,C14, F13, W13, M14}. CT Cha A, a K7 classical T Tauri star \citep{W90,GS92}, is located in the Chamaeleon I star-forming region. This region is close ($\sim160$ pc; \citealt{W97,B99,L08}) and is as young (median age $\sim2$ Myr; \citealt{L04}) as the Taurus star-forming region and IC 348. It also has relatively low extinction (typical $A_V\lesssim5$ mag; \citealt{C97}), enabling clear view of young stars. The companion CT Cha B at 2.67\arcsec (430 AU) projected separation was discovered by \cite{S08} in their VLT NACO survey. Based on its near-IR spectrum, the companion was estimated to be an M8-to-L0 ($T_{eff}\sim2600$ K) low mass ($\sim17$ $M_J$) brown dwarf with $A_V\sim5.2$ mag. \cite{S08} also imaged another closer object termed ``cc2'', whose true nature has remained puzzling \citep{S09,R12}. In this paper we present new visible AO observations providing a better measurement of $A_V$ and of the mass of CT Cha B. Our accurate astrometry allows us to determine that cc2 is, in fact, a background source.

\section{OBSERVATIONS and REDUCTION}
\begin{figure}
\includegraphics[angle=0,width=\columnwidth]{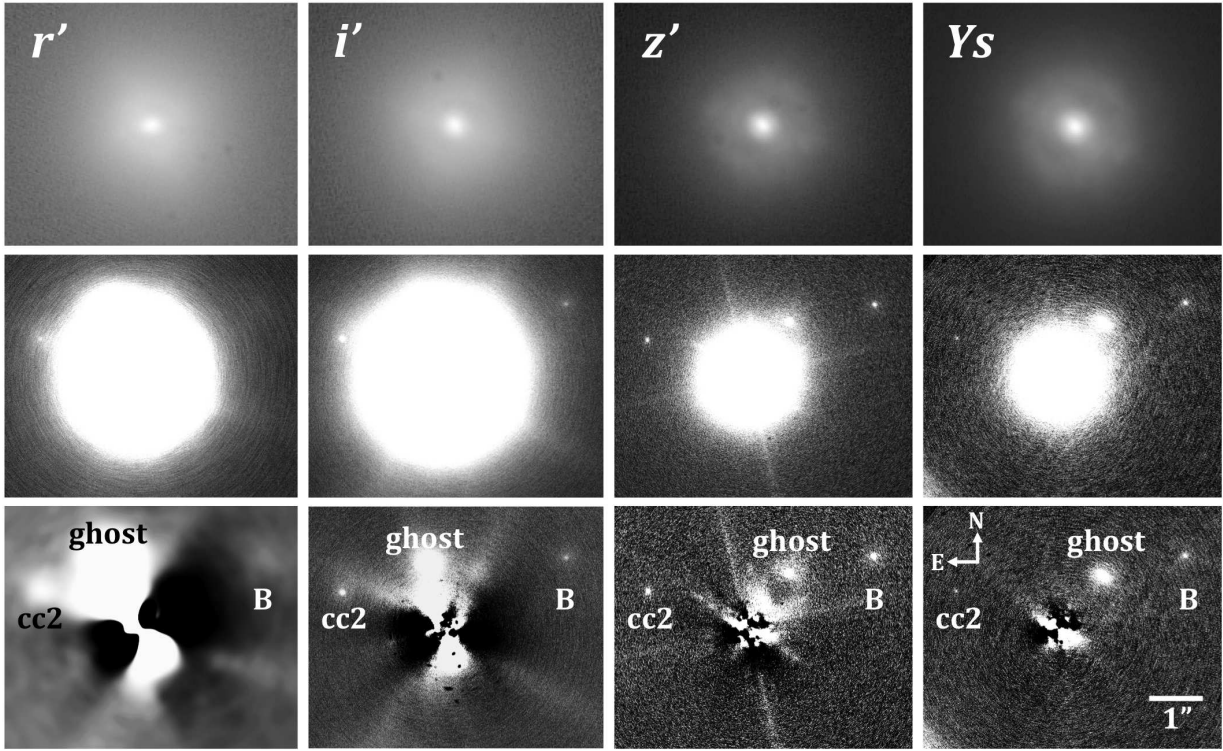}
\caption{CT Cha in MagAO filters. Image contrasts are adjusted to bring out the objects. Top row: unsaturated dataset showing the PSF. Middle row: reduced, saturated dataset before any halo subtraction. Bottom row: primary's halo removed by subtracting a rotationally symmetric PSF. The $r'$ image was further smoothed to enhance the signal-to-noise ratio. We fit the bottom row images with the DAOPHOT $allstar$ PSF fitting photometry task. At $i'$, $z'$, and $Y_S$ PSF fits using unsaturated images of A as the PSF of B were excellent, with very small ($<10\%$) PSF fitting errors. At $r'$ where B is very faint ($r'\sim22$ mag) the S/N was lower, so the fitting error increased to $>20\%$.}
\label{fig1}
\end{figure}

MagAO observations with its VisAO camera \citep{C13,M14} at $r' \,(0.62\mu$m), $i' \,(0.77\mu$m), $z' \,(0.91\mu$m), and $Y_S \,(0.98\mu$m) were performed on 2013 April 6 (UT) during the second commissioning run. Seeing was stable, ranging from 0.6\arcsec to 0.8\arcsec. We locked the AO system on CT Cha A ($R\sim12$ mag) at 100 modes and 625 Hz.\footnote{The faintness of this guide star prevented us from using all 378 modes at 1000 Hz which typically requires $R\lesssim10$ mag guide stars.} The achieved FWHMs were 0.1\arcsec, 0.08\arcsec, 0.06\arcsec, 0.06\arcsec\, for $r'$, $i'$, $z'$, $Y_S$, respectively. Strehl ratios were low due to only correcting 100 modes, since the guide star was somewhat faint for VisAO. We obtained saturated images to boost signal-to-noise ratio, with unsaturated datasets for relative photometry (top row in Figure \ref{fig1}). As a young accreting star, CT Cha A varies its brightness by $\sim0.3$ mag in the near-IR and $\sim1$ mag in the optical \citep{B98,G97,L96}. In order to calibrate the brightness of CT Cha A, we also obtained absolute photometry by observing the optical photometric standard star LTT 3864.

We carried out standard data reduction with IRAF tasks. After dark-subtraction, flat-field correction, and cross-correlation between frames, we rotated counterclockwise the saturated data by 89.11\arcdeg+parallactic angle to make north up and east left (middle row in Figure \ref{fig1}). This was following the calibration of VisAO camera based on astrometry of the Trapezium cluster \citep{C13}. We further rotated each of these images by 20, 40, ..., 340\arcdeg\, and took the median of them to approximate the halo of the primary. Then we subtracted the halo from the original images to further bring out any faint point source object (bottom row in Figure \ref{fig1}) without any loss of flux from self-subtraction. Anisoplanatic effects are still small at this small separation ($<3\arcsec$). Only our bluest filter $(r')$ showed some sign of anisoplanatism, so we smoothed the reduced $r'$ image with a gaussian (width $= 2\times\mbox{PSF FWHM}$) to enhance signal-to-noise ratio. Next we constructed a master PSF from unsaturated images of A, used it to fit CT Cha B profile in the deep images, and measured the PSF fitting flux with the DAOPHOT $allstar$ task. We note here that the on-axis CT Cha A PSF was still an excellent fit to the 2.7\arcsec\, off-axis PSF of B. Table \ref{tbl-1} summarizes our observations and PSF fitting photometry on CT Cha B. Uncertainties of B are $<0.1$ mag for $i'$, $z'$, and $Y_S$, and $\sim0.2$ mag for $r'$ due to low S/N. Near-IR ($J$, $H$, $K_S$) photometry and the $\sim 0.3$ mag uncertainty were adopted from \cite{S08}.  

To increase the accuracy of our astrometry, we also corrected for image distortion ($\lesssim15$ mas). The exact formulae to correct any residual distortions for separation $(\delta x, \delta y)$ from $(X, Y)$ are listed in \cite{C13} and reproduced here: $\mbox{true}_{\delta x} = \mbox{measured}_{\delta x} - \delta dx\times|\mbox{measured}_{\delta x}| / 110.0$ and $\mbox{true}_{\delta y} = \mbox{measured}_{\delta y} - \delta dy\times|\mbox{measured}_{\delta y}| / 44.5$, where $\delta dx = -0.00038921676 \times (X-512)+0.00084322443 \times (Y-512)$ and $\delta dy = -0.00025760395\times(X-512)-0.0024045175\times(Y-512)$. We also retrieved archive HST and VLT NACO data for proper motion analysis. The HST data were already reduced by the OPUS pipeline and the MultiDrizzle software, as described in \cite{R12}. We reduced NACO raw frames by shift-and-add, without flat-field correction and dark-subtraction. NACO image distortion is very small and only up to 3 mas at field edges \citep{N08}. The error budget of our measurements includes platescale and centroid uncertainties. Table \ref{tbl-2} lists our astrometric measurements.

\section{RESULTS}
\subsection{Optical Images}
Figure \ref{fig1} shows the CT Cha system seen with our broad-band filters, with CT Cha B and cc2 visible in all four. This is the first optical detection of CT Cha B, as it was not detected in previous HST narrow-band optical observations \citep{R12}. Judging from its color, cc2 is relatively blue ($r'-i'=0.9$) so unlikely to be another low-mass companion. \cite{R12} also speculated that a faint ``object'' seen at [OI] $1.5\arcsec \,$to the south of CT Cha A could be real, but we cannot confirm any other faint object in our images, especially at $r'$ where a narrow band [OI] or H$\alpha$ source might have been visible. Thus, it is unlikely to be a real object.

\subsection{Astrometry}
\begin{figure}
\begin{tabular}{cc}
\includegraphics[angle=0,width=\columnwidth]{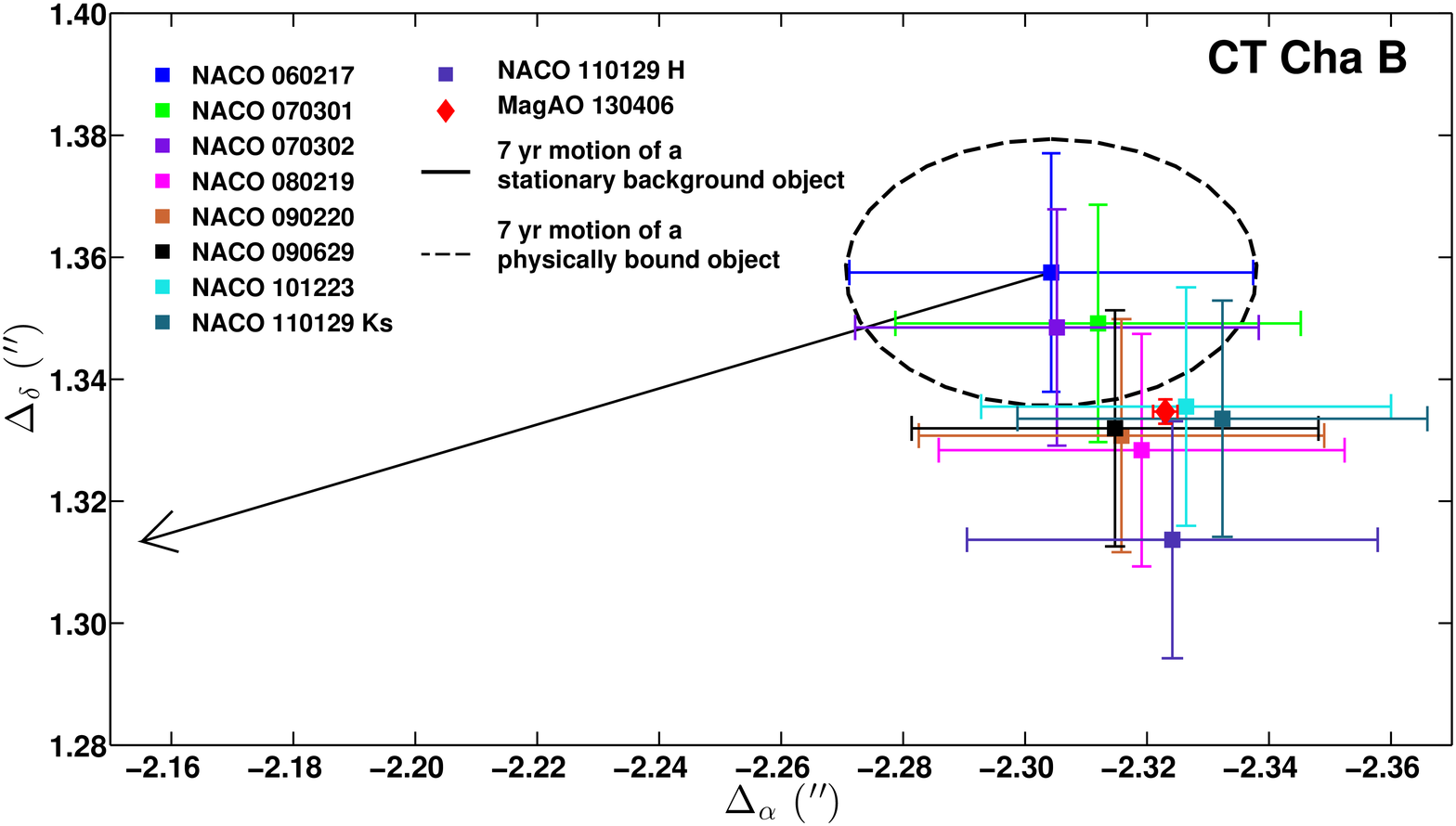}\\ 
\includegraphics[angle=0,width=\columnwidth]{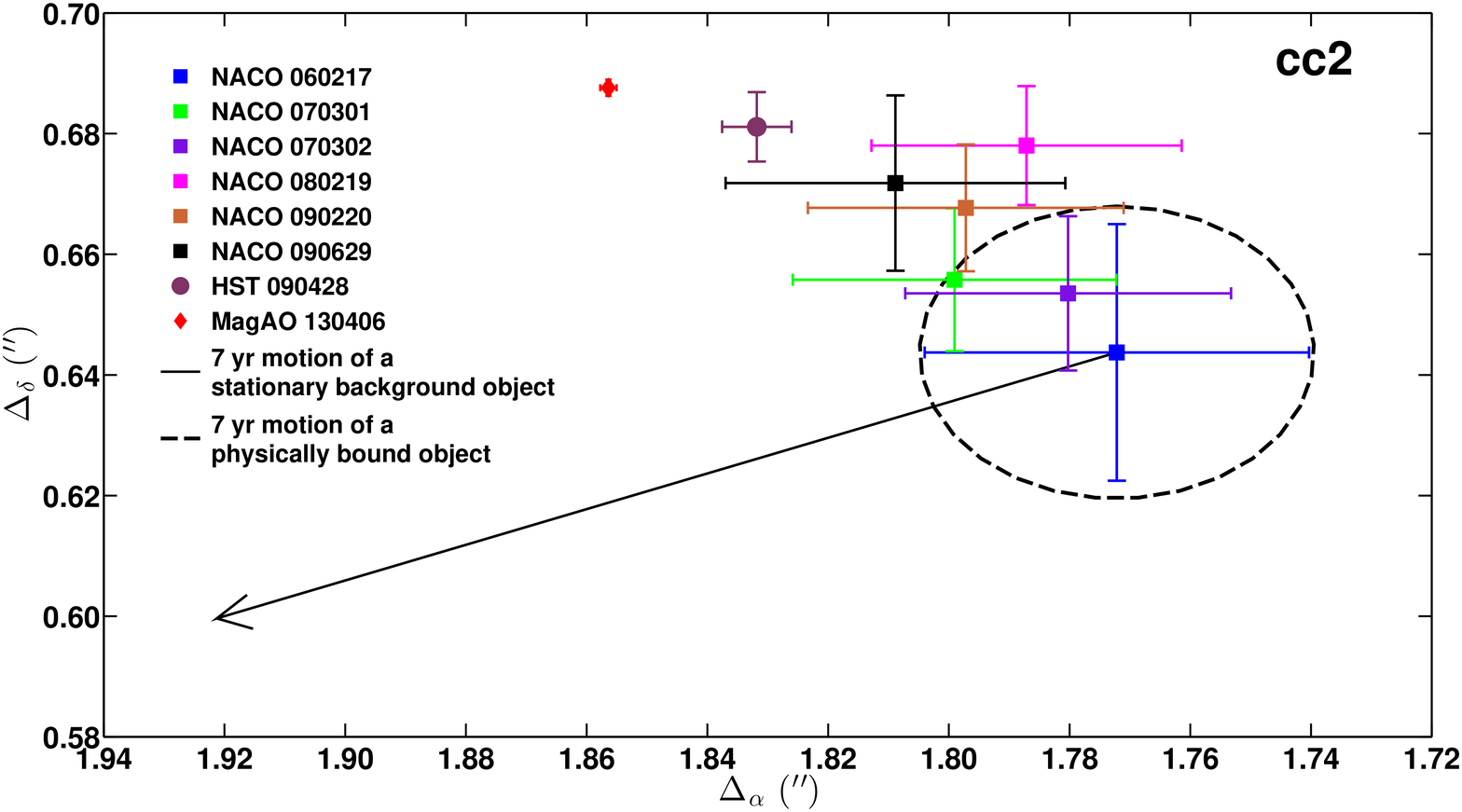}
\end{tabular}
\caption{Seven-year relative astrometry for CT Cha B (top) and cc2 (bottom). Proper motion of CT Cha A ($\mu_\alpha \cos\delta=-21.3$ mas/yr, $\mu_\delta=6.3$ mas/yr) is from \cite{S08}. The ellipse in dashed line shows the convolution between the orbital motion and astrometric uncertainty. Apparently cc2 is not a bounded nor a non-moving background object. It is, in fact, a background star moving northwestward at $\sim15.4$ mas/yr ($\mu_\alpha \cos\delta\sim-8.2$ mas/yr and $\mu_\delta\sim13.0$ mas/yr;  absolute proper motion). For CT Cha B, the MagAO value is consistent with those measured from archive NACO images, showing that it is a physical companion.}
\label{fig2}
\end{figure}

The nature of cc2 is not fully settled in literature. \cite{S09} presented two-year astrometry, showing that it is likely to be a background star. But \cite{R12} suggested that cc2 may be physically associated based on their single epoch HST observations. We measured the positions of cc2 and CT Cha B in images taken by various instruments over $\sim7$ year time span (Figure \ref{fig2}). Significant common-proper motion has been found for CT Cha B, confirming it is physically bound. However, we detected a significant non-common $\sim15.4$ mas/yr northwestward motion for cc2, unambiguously demonstrating that it is not a co-moving companion but instead a background star, and not a member of Chamaeleon I.

\subsection{SED Fitting and Derived Properties}
\subsubsection{Effective Temperature}
\begin{figure}
\includegraphics[angle=0,width=\columnwidth]{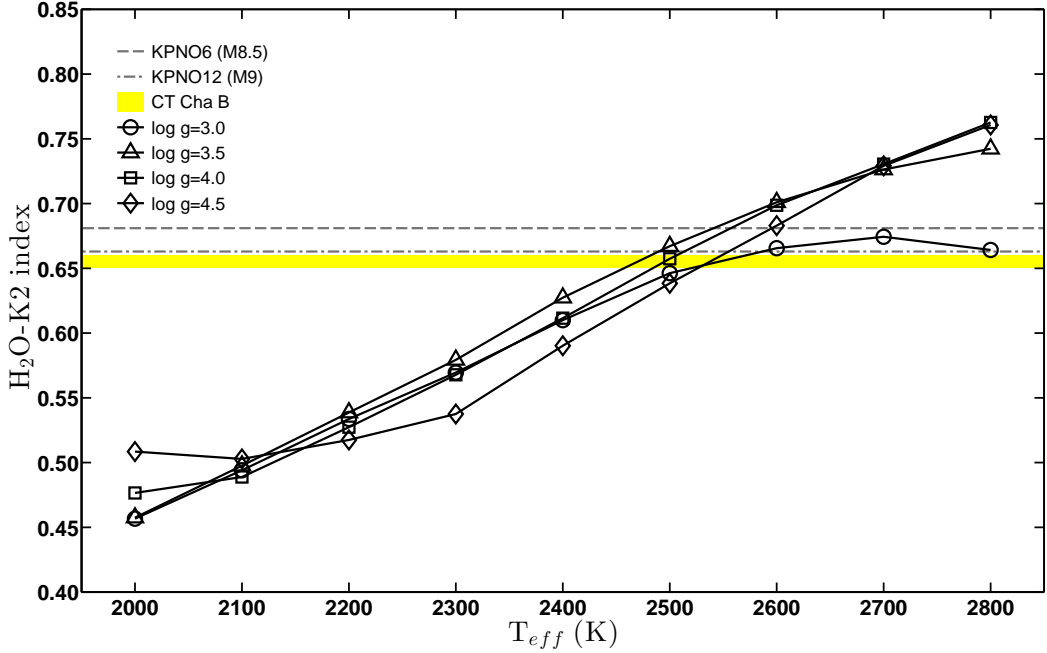}
\caption{Variation of the ${\mbox{H}_2}$O-K2 index with surface gravity for $2000-2800$ K BT-Settl synthetic spectra. We found that for log $g=3.0$ to 4.5 the models give similar indices. Hence this is a reasonably gravity insensitive $T_{eff}$ index. CT Cha B's index is similar to that of other young late M-dwarfs in the Taurus star-forming region \citep{M07}, and is more consistent with the 2500 K models. We conservatively pick $2500\pm100$ K for CT Cha B.}
\label{fig3}
\end{figure}

To further narrow down the uncertainty of $T_{eff}$, we retrieved the spectrum taken with the VLT SINFONI spectrograph \citep{S08}, and calculated the ${\mbox{H}_2}$O-K2 index following the prescription of \cite{RA12}. Assuming solar metallicity, we found that for CT Cha B this index is almost independent of extinction, ranging from 0.65 to 0.66 for $A_V=0$ to 5.5 mag. In Figure \ref{fig3}, we plotted the variation of the index with a range of surface gravity for $2000-2800$ K BT-Settl atmospheric models \citep{A11}. Within this temperature range, ${\mbox{H}_2}$O-K2 index is rather insensitive to log $g$. Hence we are free to use it for young cool objects like CT Cha B. Our best fit to the index corresponds to a spectral type M$9\pm1$ with $T_{eff}=2500\pm100$ K. 

\subsubsection{Extinction, Bolometric Luminosity, and Mass}
CT Cha B was previously estimated from near-IR spectroscopy to have an extinction higher than its host star ($A_V=$5.2 mag versus 1.3 mag; \citealt{S08}). Our data benefit from the fact that visible wavelengths are more sensitive to dust extinction, so we can determine $A_V$ at higher precision with MagAO's VisAO camera.

\begin{figure}
\includegraphics[angle=0,width=\columnwidth]{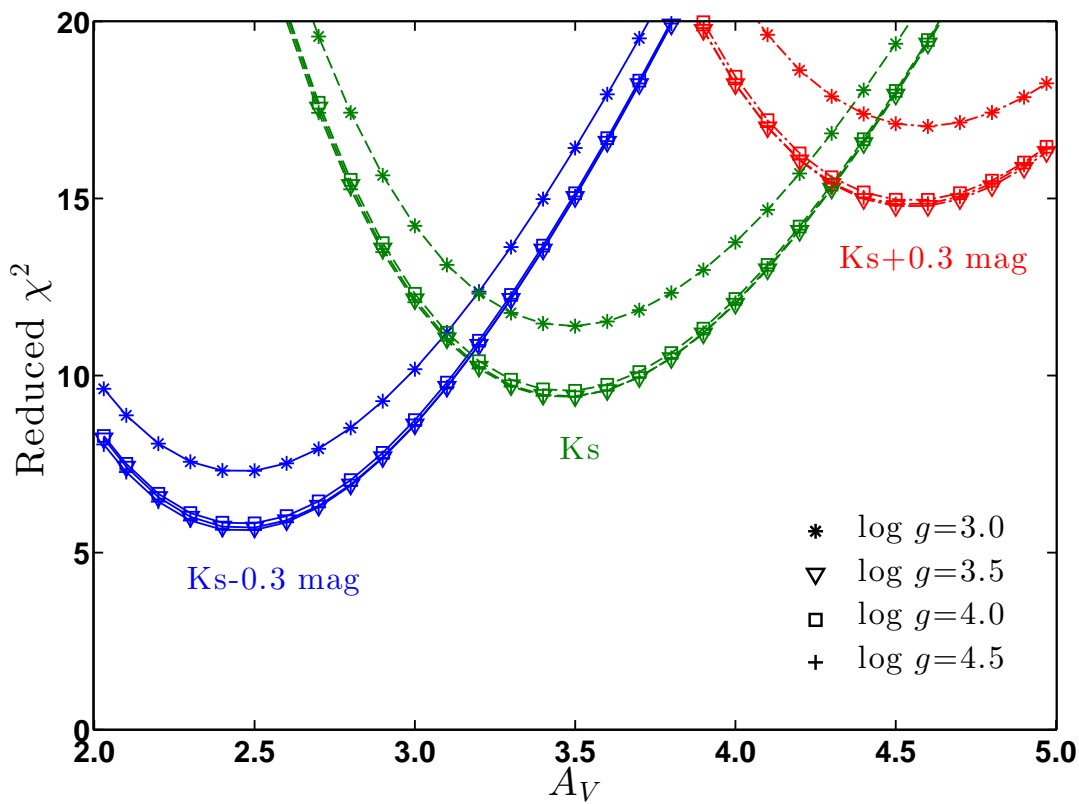}
\caption{Reduced $\chi^2$ as a function of extinction for $\pm0.3$ mag normalization uncertainty and different surface gravity. Note that the degree of freedom (fitting photometry at $i'$, $z'$, $Y_S$, $J$, and $H$) is 3 because we normalized the model at $K_S$. If we fit the minima of these curves, then this analysis favors $A_V=3.4\pm1.1$ mag, independent of surface gravity. The relatively high $\chi^2_r$ is likely due to some systematic errors that are discussed in section 3.3.2 of the text.}
\label{fig4}
\end{figure}

\begin{figure}
\includegraphics[angle=0,width=\columnwidth]{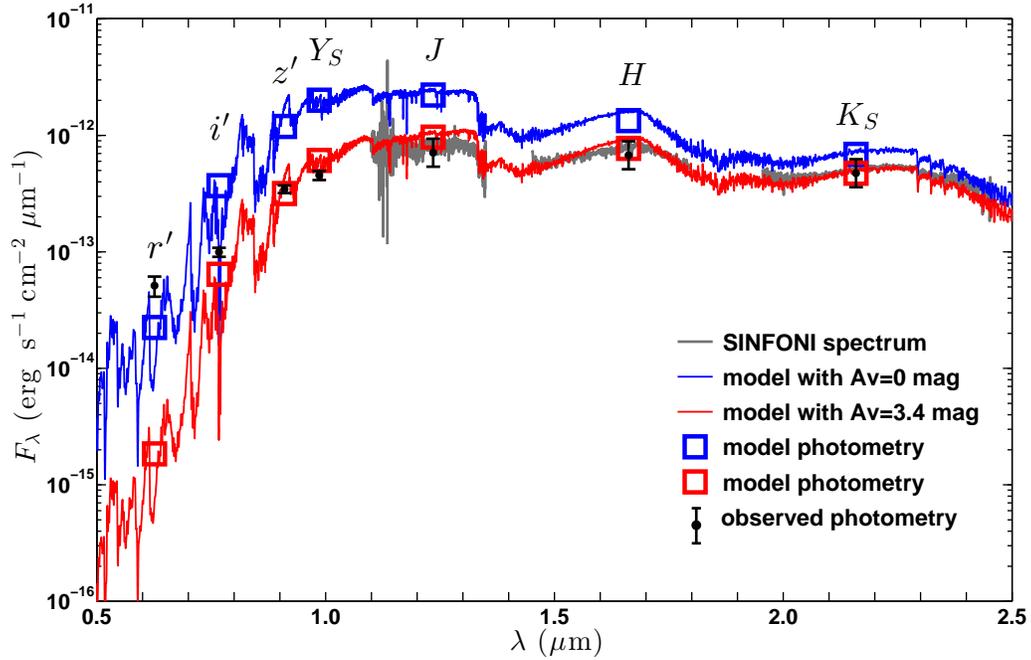}
\caption{SED fitting with the reddened 2500 K BT-Settl spectrum. The $JHK_S$ photometry was adopted from \cite{S08}, and the extinction law was from \cite{F99}. We matched the reddened model with the observed $K_S$ flux, and calculated the reduced $\chi^2$ to determine the extinction (see Figure \ref{fig4}). We only utilized the apparent fluxes in order to avoid the $10-20\%$ uncertainty in CT Cha's distance. The $r'$ photometry is very much higher than photospheric due to H$\alpha$ accretion luminosity. Note that $r'$ was not used in fit of the $\chi^2_r$ due to H$\alpha$ emission.}
\label{fig5}
\end{figure}

We applied multiple values of $A_V$ following the extinction law in \cite{F99} to redden the 2500 K BT-Settl synthetic spectra normalized at $K_S$ (Figure \ref{fig4}). Minimization of the reduced $\chi^2$ is based on the reddened models fit to the observed $i'$, $z'$, $Y_S$, $J$, and $H$ photometry (black points in Figure \ref{fig5}). We found that while the result is independent of surface gravity, $\chi^2_r$ remains high even after including the $\pm0.3$ mag uncertainty at $K_S$. Some systematic errors may come into play. For example, the adopted extinction law might be invalid due to grain growth in the disk, or there could be multiple dust components. On the other hand, scattered light from the disk or outflow gas may contribute to our $i'$ photometry, as in the case of R Mon \citep{Close97}. In this picture, blue light follows indirect paths to the observer, avoiding passing through the disk and making our extinction estimate likely a lower limit. Another possible cause for higher $\chi^2_r$ is that an overall offset $\sim0.5$ mag might exist between the visible and near-IR data because they were taken on different nights and CT Cha A is a well-known variable. Finally, the companion itself could also be variable in the visible just like the primary due to accretion. In any case, with no prior knowledge of the material around CT Cha B, our current data yield a best fit to a lower extinction $A_V=3.4\pm1.1$ mag. We plotted the reddened synthetic spectrum together with the observed photometry in Figure \ref{fig5}.

\begin{figure}
\includegraphics[angle=0,width=\columnwidth]{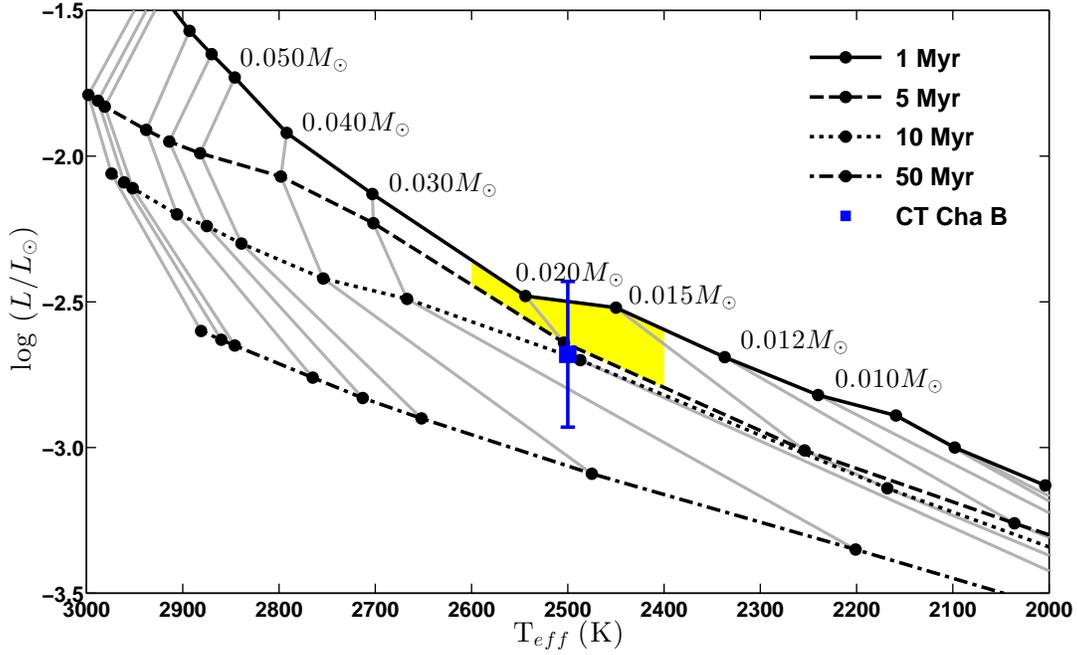}
\caption{Color-Magnitude diagram with DUSTY evolutionary tracks. The shaded area is the only area consistent with the observed $T_{eff}$ (2400--2600 K), luminosity (log$(L_{bol}/\Lsol)=-2.68\pm0.25$), and estimated age of CT Cha B (1--5 Myr). This yields a mass estimate of $\sim 14-24$ $M_J$, consistent with \cite{S08}. CT Cha B is very unlikely to be under the deuterium-burning limit, so it is a very low-mass brown dwarf.}
\label{fig6}
\end{figure}

We followed the approach in \cite{H97} to calculate the bolometric luminosity. We converted our $i'$ photometry to Cousins $I_C$, de-reddened it by $A_V=3.4$ mag, applied the bolometric correction from \cite{T93} and \cite{B95}, and obtained log$(L_{bol}/\Lsol)=-2.68\pm0.25$. As a comparison, we also calculated $L_{bol}$ from the $K$ flux following \cite{C07} and had a similar value log$(L_{bol}/\Lsol)=-2.71\pm0.20$. Both values are consistent with log$(L_{bol}/\Lsol)=-2.68\pm0.21$ in \cite{S08}. We also calculated CT Cha B's radius using $L\propto R^2T^4$ and obtained $\sim 2.4 R_{J}$. Then we applied the DUSTY evolutionary tracks \citep{C00,B01} to derive a mass estimate of $\sim 14-24 M_J$ based on the $\sim1-5$ Myr age and $T_{eff}$ (Figure \ref{fig6}). Therefore, CT Cha B is most likely a very low mass brown dwarf, just above the planetary mass limit.

\subsubsection{Accretion Rate}
Pa$-\beta$ emission, an accretion signature, has been seen in CT Cha B's $J-$band spectrum \citep{S08}. Since CT Cha B is widely separated from the host star, it may harbor its own disk and still be actively accreting at this time. Figure \ref{fig5} shows that our $r'$ detection is about 20 times brighter than its predicted continuum. This significant $r'$ excess seems to imply strong H$\alpha$ emission from accretion, allowing us to calculate the mass accretion rate. Attributing $>95\%$ of the $r'$ flux to H$\alpha$ and following the prescription of \cite{C14}, we estimated $\dot{M}\sim6\times10^{-10}\Msol/\mbox{yr}$, which is reasonable as it implies that a few $M_{J}$ of brown dwarf mass could be accreted in a few million years at the end of the gas-rich disk phase. The accretion rate we derived is also consistent with recent HST observations by \cite{Z14}, who measured $\dot{M}\sim10^{-11}-10^{-9} \Msol/\mbox{yr}$ for three substellar companions GSC 6214-210 B, GQ Lup B, and DH Tau b based on their optical excess.

\subsection{Implications}
The different extinction between the primary and the secondary may imply that both objects have their own disks likely with different inclination angles, resembling conceptually the configuration of HK Tau A and B \citep{JA14}. CT Cha B's $r'$ excess, together with other accreting objects in \cite{Z14}, suggest that accretion disks could be common among young low-mass companions and favor the ``star-like'' formation via gravitational collapse and fragmentation of molecular clouds. The survival of these significant disks also implies that substellar companions form near their current locations rather than being ejected there \citep{K14}. Strategic H$\alpha$ surveys such as MagAO's ongoing Giant Accreting Proto-planets Survey (GAPplanetS) may have the potential to probe $\sim 1 M_{J}$ accreting giant planets and shed light on the earliest stage of planet formation \citep{C14}.

\section{Summary}
MagAO observations on CT Cha at $r'$, $i'$, $z'$, and $Y_S$ have improved the accuracy of the extinction towards CT Cha B. The companion is detected in all of our optical filters, whereas no detections were made by HST. It is over-luminous at $r'$, indicating active accretion at a rate of $\dot{M}\sim6\times10^{-10}\Msol/\mbox{yr}$. The ${\mbox{H}_2}$O-K2 index derived from the $K_S$ spectrum is consistent with a $T_{eff}=2400-2600$ K brown dwarf. Using the BT-Settl model, we show that CT Cha B is best fit by $A_V=3.4\pm1.1$ mag, which is lower than previous estimate and translates to a mass estimate of $14-24$ $M_J$ based on the DUSTY tracks. We do not see the faint southern [OI] source seen in previous HST observations, so it is unlikely to be real. Finally, our astrometry on cc2 is incompatible with a previous claim that it is a co-moving object.

\acknowledgements
We thank the anonymous referee for helpful comments that greatly improved this manuscript. We thank Professor Jennifer Patience for kindly providing the SINFONI spectra of CT Cha B. We thank the whole Magellan Staff for making this wonderful telescope possible. We would especially like to thank Povilas Palunas (for help over the entire MagAO commissioning run). Juan Gallardo, Patricio Jones, Emilio Cerda, Felipe Sanchez, Gabriel Martin, Maurico Navarrete, Jorge Bravo and the whole team of technical experts helped do many exacting tasks in a very professional manner. Glenn Eychaner, David Osip and Frank Perez all gave expert support which was fantastic. It is a privilege to be able to commission an AO system on such a fine telescope and site. The MagAO system was developed with support from the NSF MRI and TSIP programs. The VisAO camera was developed with help from the NSF ATI program. YLW's and LMC's research were supported by NSF AAG and NASA Origins of Solar Systems grants. JRM is grateful for the generous support of the Phoenix ARCS Foundation. JRM and KM were supported under contract with the California Institute of Technology funded by NASA through the Sagan Fellowship Program.

\clearpage
\begin{deluxetable}{ccrccccccc}
\tabletypesize{\normalsize}
\tablecaption{MagAO Photometry on CT Cha B\label{tbl-1}}
\tablewidth{0pt}
\tablehead{
\colhead{Filter} &
\colhead{$t_{sat}$} &
\colhead{$t_{unsat}$} & 
\colhead{$m$} &
\colhead{$\Delta m$\tablenotemark{a}} &
\colhead{$F_\lambda$\tablenotemark{b}} &\\
 & 
\colhead{(s$\times\#$)}&
\colhead{(s$\times\#$)}&
\colhead{(mag)}&
\colhead{(mag)}&
\colhead{($10^{-13}$ erg $\mbox{s}^{-1} \mbox{cm}^{-2} \mu \mbox{m}^{-1}$)}&
}
\startdata
$r'$    & $20\times75$    &  $2.27\times13$   & $21.71\pm0.21$ & 9.80 & $0.51 \pm 0.10$ \\          
$i'$    & $30\times52$    &  $2.27\times14$      &  $20.32\pm0.09$   &   9.22   & $1.00\pm0.09$ \\           
$z'$   &   $15\times13$   &  $2.27\times13$     &  $18.46\pm0.08$   &   7.96   &  $3.46\pm0.26$ \\
$Y_S$  & ...		     &  $30\times35$        &  $17.94\pm0.09$   &   7.59   &  $4.55\pm0.41$ \\
\enddata
\tablenotetext{a}{Relative to CT Cha A, measured with PSF fitting photometry.}
\tablenotetext{b}{Calibrated with the standard LTT 3864 taken at a similar air mass to CT Cha in photometric conditions (errors are $<1\%$ due to different air masses from model atmosphere). We used photometry calibrations of \cite{C13} and \cite{M14}.}
\end{deluxetable}

\clearpage
\begin{deluxetable}{ccccr}
\tabletypesize{\normalsize}
\tablecaption{MagAO Astrometry on 2013 April 6 (UT)\label{tbl-2}}
\tablewidth{0pt}
\tablehead{
\colhead{Filter} &
\colhead{Plate Scale (\arcsec/pixel)} &
\colhead{Name} &
\colhead{Separation (\arcsec)} &
\colhead{P.A. (\arcdeg)} }
\startdata
\multirow{2}*{$r'$}   &  \multirow{2}*{$0.007917\pm0.000015$}		&    B       & $2.717\pm0.030$ & $298.8\pm0.6$ \\
        			      & 	&  cc2       &     $1.962\pm0.004$   &    $69.8\pm0.3$    \\  [6pt]        
\multirow{2}*{$i'$}   &   \multirow{2}*{$0.007907\pm0.000015$}		&    B        &     $2.671\pm0.006$   &    $300.0\pm0.3$   \\
                               &		&   cc2      &     $1.962\pm0.004$   &    $69.7\pm0.3$     \\   [6pt]      
\multirow{2}*{$z'$}  &  \multirow{2}*{$0.007911\pm0.000012$}		&    B        &     $2.679\pm0.004$   &    $300.0\pm0.3$   \\
                               &		&  cc2	&     $1.995\pm0.003$   &    $69.7\pm0.3$      \\	 [6pt]          
\multirow{2}*{$Y_S$}  & \multirow{2}*{$0.007906\pm0.000014$}	&    B        &     $2.684\pm0.005$   &    $299.9\pm0.3$    \\
                                    &		&  cc2      &     $1.991\pm0.005$   &    $69.6\pm0.3$
\enddata
\end{deluxetable}

\clearpage

\clearpage

\clearpage

\clearpage

\clearpage

\clearpage

\end{document}